*Article*

# Big Data Analytics and Its Applications

**Mashooque A. Memon[1,*], Safeeullah Soomro[2], Awais K. Jumani[3] and Muneer A. Kartio[3]**

[1]Department of Computer Science, Benazir Bhutto Shaheed University, Layari, Karachi
pashamorai786@gmail.com
[2]Department of Computer Science, AMA International University, Bahrain
s.soomro@amaiu.edu.bh
[3]Department of Computer Science, Shah Abdul Latif University, Khairpur Mirs Pakistan
awaisjumani@yahoo.com; kartiomuneer@gmail.com
**\***Correspondence: pashamorai786@gmail.com

**Abstract:** The term, Big Data, has been authored to refer to the extensive heave of data that can't be managed by traditional data handling methods or techniques. The field of Big Data plays an indispensable role in various fields, such as agriculture, banking, data mining, education, chemistry, finance, cloud computing, marketing, health care stocks. Big data analytics is the method for looking at big data to reveal hidden patterns, incomprehensible relationship and other important data that can be utilize to resolve on enhanced decisions. There has been a perpetually expanding interest for big data because of its fast development and since it covers different areas of applications. Apache Hadoop open source technology created in Java and keeps running on Linux working framework was used. The primary commitment of this exploration is to display an effective and free solution for big data application in a distributed environment, with its advantages and indicating its easy use. Later on, there emerge to be a required for an analytical review of new developments in the big data technology. Healthcare is one of the best concerns of the world. Big data in healthcare imply to electronic health data sets that are identified with patient healthcare and prosperity. Data in the healthcare area is developing past managing limit of the healthcare associations and is relied upon to increment fundamentally in the coming years.

**1. Introduction**

In order to explain a new model for data utilization, the locution "Big Data" has recently been emerging. These new hi-techs, in the field of Information Technology, tend to emerge very often and with a huge publicity, however the point of difference takes some time to be recognized. Big Data (also known as BD) is distinct in numerous ways such as volume (too big), velocity (faster arrival) variability (quick changes), veracity (much commotion), and variety (diversity). Using orthodox propositions and procedures this Big Data is processed in limited reckoning arrangements. Even the technologies introduced to support BD contain different variety of presentations, which ultimately make it hard to stimulate the creation of tools and applications to help encompass data from numerous sources. This study therefore identifies possible areas for uniformity within the BD technology expanse [2].

Multifaceted and huge datasets have various types of different and important features that are closely in resemblance with "Big Data". To administer these datasets is troublesome with the traditional information preparing frameworks. Furthermore, data storage, data transition, data visualization, data penetrating, data analysis, data security, data privacy violations and sharing propose different uphill challenges that the "Big Data" reinforces [1].

To potentially grasp the supplementary information sets the Big Data is appropriate as a contemplation that highlights the dearth of ability of habitual information structures. The emergence of Big Data model transpires when the compute of the data is either in take it easy or in motion; it forcedly induces the management of data in the system engineering design to become a significant driver. Basically the Big Data Model represents a paradigm shift in the data infrastructures i.e. from substantial systems with perpendicular mount into a parallel mounted system that coalesce an unbounded connected set of reserves. This change from perpendicular to parallel predicates some different problems in some dissimilar areas such as information deliverance, information orchestrating, and inactivity in the consistency across schematics, stack stabilizing, and process deficiencies and their interdependencies on single hand. On other hand, the Big Data model uses different contraptions to provide the clamber in data handling, but embodies the same shift again. The reason for this move is to bargain out codes and information crosswise over inexactly coupled assets and match the scaling in information. As to produce additional knowledge about the data a different purpose of residing and retrieve huge amounts of data is to execute analysis. In the olden days, the assay was usually attained on an undirected sample of the data [2].

The word "Big Data" contains assortment of distinctiveness, it is used in various contexts. To identify with where ideology will appropriately assist backing the big data model, in order to find what the term really means we have to stretch our knowledge to some extent of consonance. The "Big Data" is a gathering of information with unique excellence (e.g. capacity, momentum, array, range, precision, etc.) that for a problem realm at any given moment can't be expertly handled using current/accessible/apperceived/routine advancements and strategies with a specific end goal to concentrate esteem.

The above definition recognizes of Big Data from business knowledge and conventional value-based administration while suggesting an expansive range of utilizations that incorporates them. A definitive objective of handling Big Data is to get separated esteem that can be trusted (in light of the fact that the central information can be trusted) [2].

This is done through the usage of higher examination next to the entire measure of data disregarding scale. Parsing these target edges the regard exchange for Enormous Information utilizes cases.

Figure 1 demonstrates a practical view of Big Data.

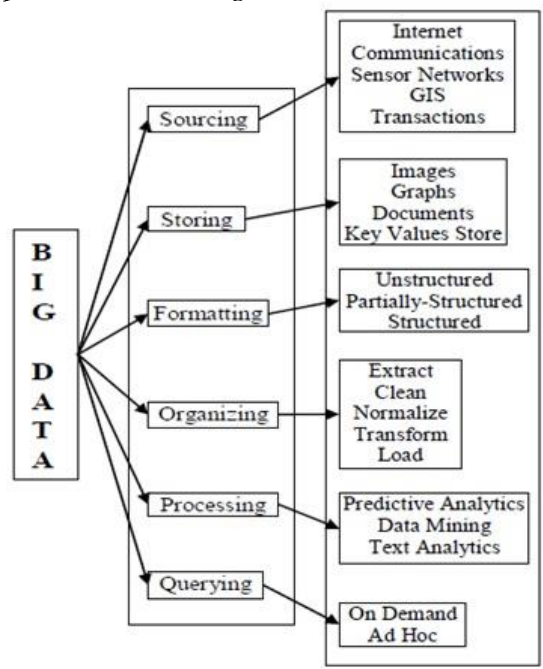

**Figure 1**. Illustrates the Practical View of Big Data

- Sourcing: Data sources distinguished are Internet Web Pages, Discussion Forums, Chats and message made in and among informal social organizations or networks, Remote Sensing Networks, All sorts of everyday exchanges done through web based applications.
- Designs: Unstructured, in part organized, and organized

- Storing: Pictorial based, Graphic based, written archives, Key Value Stores (Key Values Store is a method for putting away application's information with invalid blueprint. It doesn't require a static information show. Remarkable keys are utilized to speak to values put away in it.) Storing: Image based, Graph based, reports, Key Value Stores (Key Values Store is a method for putting away application's information with invalid mapping. It doesn't require a static information demonstrate. Interesting keys are utilized to speak to values put away in it.)
- Organization: Withdraw, Clear, Standardize, Change, Load.
- Handle: Online, Offline
- Enquiry: On request, impromptu

## 2. Data Analytics

The Big Data Analytics is a strategy used to analyze colossal information sets containing assorted qualities of information sorts, for example, enormous information to reveal every single shrouded example, obscure relationship, advertise drift, client inclinations and other supportive business data. These demonstrative outcomes could prompt to proficient advertising, new income openings, enhanced client benefit, enhanced operational skill, and upper hands over contender associations and different business repayment [3].

Big Data has turned into a center theme in various ventures and research teaches and in addition for society as a whole. This is on the grounds that the capacity to create, gather, convey, prepare and examine exceptional measures of differing information has practically widespread utility and serves to essentially change the way businesses work, how research should be possible and how individuals live and utilize present day innovation. Distinctive ventures, for example, car, fund, social insurance or assembling, can significantly profit by enhanced and speedier information investigation, e.g., as represented by current industry patterns like "Industry 4.0" and "Web of Things". Information driven research approaches using Big Data innovation and investigation turn out to be progressively ordinary, e.g., in the life sciences, geo sciences or in cosmology. Clients using PDAs, web-based social networking, and web assets invest expanding measures of energy on the web, create and expend colossal measures of information and are the objective for customized administrations, proposals and notices. A large portion of the conceivable advancements identified with Big Data are still in an early stage yet there is awesome guarantee if the differing mechanical and application-particular difficulties in overseeing und utilizing Big Data are effectively tended to. A portion of the specialized difficulties have been related to various "V" attributes, specifically Volume (support of very high data volumes), Velocity (fast analysis of data streams), Variety (support for diverse kinds of data) and Veracity (support for high data quality). Different challenges identify with the assurance of individual and touchy information to guarantee a high a level of protection and the capacity to transform the tremendous measure of information into valuable bits of knowledge or enhanced operation.

- Analytics can be classified in to three types they are: Predictive Analytics, Descriptive Analytics and Prescriptive analytics [9].
- Descriptive analytics: The most straightforward class of investigation," one that permits you to consolidate huge information into littler, more valuable chunks of data.
- Predictive analytics: It is the following stride up in reduction of data. It uses an assortment of measurable, displaying, information mining, and machine learning strategies to study later and verifiable information, along these lines permitting experts to make forecasts about what's to come.
- Prescriptive analytics: It is a type of predictive analytics. It's essentially when we have to endorse an activity, so the business decision maker can take this data and act.

## 3. Big Data Application

Big Data is nearly ubiquitous. Every business such as health or general living standards could apply big data analytics. Big data is a field which can be used in any area whatsoever given that this big quantity of data can be harnessed to one's advantage. The major applications of big data are listed below [4].

The Third Eye-Data idea: Organizations globally are recognizing the importance of big data analytics. Big Data analytics is one stop solution for almost every organization, it helps predict customer purchasing behavior pattern, detecting fraud and abuses. It provides opportunity to business experts to questions and understands data according to their business need irrespective of difficulty and volume of the data. This can be accomplished by professionally visualizing and presenting the data in an understandable manner. Giants like Google, Facebook, Twitter, eBay, Wal-Mart etc., also adopt data visualization to effortlessness difficulty of handling data. Data visualization has shown enormous positive outcomes in such business organizations. Implementing data analytics and data visualization; enterprises can lastly begin to tap into the huge potential that big data possesses and make sure greater return on investments and business constancy.

In Banking: The make use of customer data could also move up privacy issues. By spotting inconspicuous associations between it appears that unmistakable bits of information, Big Data Analytics could possibly uncover sensitive individual data. A look into demonstrates that 62% of financiers are careful in their utilization of Big Data because of isolation issues [3]. Further, outsourcing of information investigation execution or sharing of client information crossways offices for the production of happier understanding additionally open up security dangers.

In Agriculture: A biotechnology firm uses sensor data to enhance procuring efficiencies. It plants examination collects and runs reenactments to figure how plants respond to various changes in condition. Its information surroundings over and over acclimates to change in the quality of different information it gathers, and temperature, water levels, soil arrangement, development, yield, and quality sequencing of every plant in the proving ground. These recreations approve to locate the ideal ecological circumstance for correct quality sorts.

In Finance: Finance related organizations are utilizing outsider acclaim scoring while assessing new acclaim applications. Be that as it may, the banks are at present utilizing their own acclaim scoring investigation for accessible clients utilizing an expansive scope of information, and also checking, reserve funds, charge cards, home loans, and venture information.

In Economy: Expected starting from the earliest stage to bargain brilliantly with ware equipment, Hadoop can assist associations for move to low budget servers.

**4. Big Data Analytics Tools & Techniques**

Within the background of Big Data is intrinsic in increasing the volume of data and require for high accessibility information, however, straight architectures physical constraints compulsory by the topology attached to them. It is a bendy and highly available architecture for big scale computation and data processing on a network of commodity hardware. To fill up this gap appeared Apache Hadoop, a framework for processing and storage data in large-scale. Hadoop is an Apache open source structure written in java that permits scattered preparing of datasets across over network of PCs utilizing simple programming models. The Hadoop venture was made in 2005 by Doug Cutting, who put the name of Hadoop was out of appreciation for his child, as this was his child's teddy bear's name. Doug Cutting built up a structure of circulated documents in light of papers gave by Google on Map Reduce (prematurely ended) and GFS (Google File System), not long after the venture was Yahoo's speculation confronted issues to handle the vast number of references for sites, from that point Hadoop appeared as a free venture of the Apache Software Foundation. In January 2008, Hadoop has become a high significance of the Apache project, confirming its achievement and its varied active community. Now, Hadoop was being utilized by numerous different organizations not withstanding Yahoo, as Last.fm, Facebook, and the New York Times (WHITE, TOM, 2013). Apache Hadoop has been completed in Java and has its open source code. Additionally bolsters the capacity and access to the expansive volume of information, which may physically be on a solitary PC or even on various PCs, which are alluded to us. Running on Linux environment the fundamental target of Hadoop is to store huge scale data in a dispersed domain empowering fast get to and conquer the significant difficulties of Big Data.

The use of Apache Hadoop apply the thought of master/slave in a network environment of computers connected by an Ethernet network, where computers, also called nodes within a cluster have dissimilar functions. The master and slave nodes work in sync, allowing the       operation of Hadoop. Apache Hadoop is intended to be implemented at any level of clusters, whether it consists of a network of dispersed high-end servers or even on a home network of personal computers. Utilizing the capacities of Apache Hadoop is conceivable to coordinate a few databases, whether social, non-social, organized or unstructured. The structure of Hadoop allows you to store and incorporate all these diversity. The Apache Hadoop system includes not just the idea of Big Data and database, additionally disseminated computing, parallel processing, document frameworks, working frameworks, differences of data (organized and unstructured information) and record taking care of through its own document framework HDFS (Hadoop File System), permitting through its different components bring together the operation through the structure.

Framework of tools which concerned with Hadoop is depicted in Figure 1.

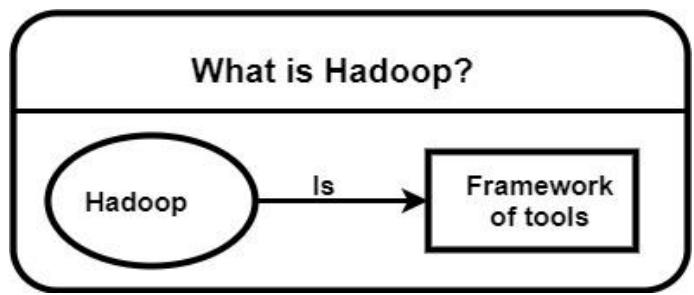

**Figure 2**. Represents the Hadoop

The objectives of running applications on Big Data shows the supports on Hadoop which is illustrates in Figure 3.

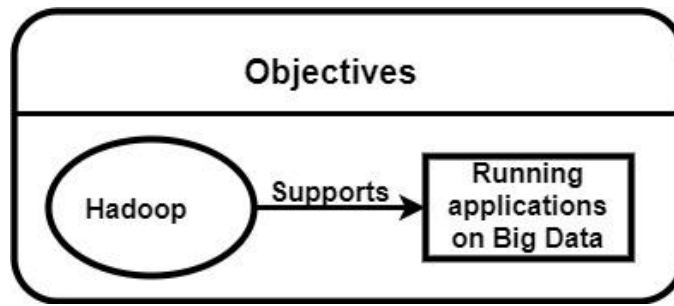

**Figure 3**. Shows the Running applications on Big Data

The open source platform which is related to Hadoop is apache and that is shows in Figure 4.

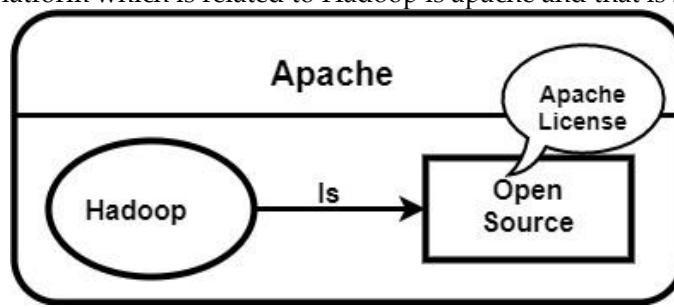

**Figure 4**. Illustrates the Apache License

**5. Hadoop Framework Architecture**

Hadoop framework consists of two main core components
- Distributed file system (HDFS)
- Execution engine (MapReduce)

Hadoop Master/Slave Architecture is depicted in Figure 5.

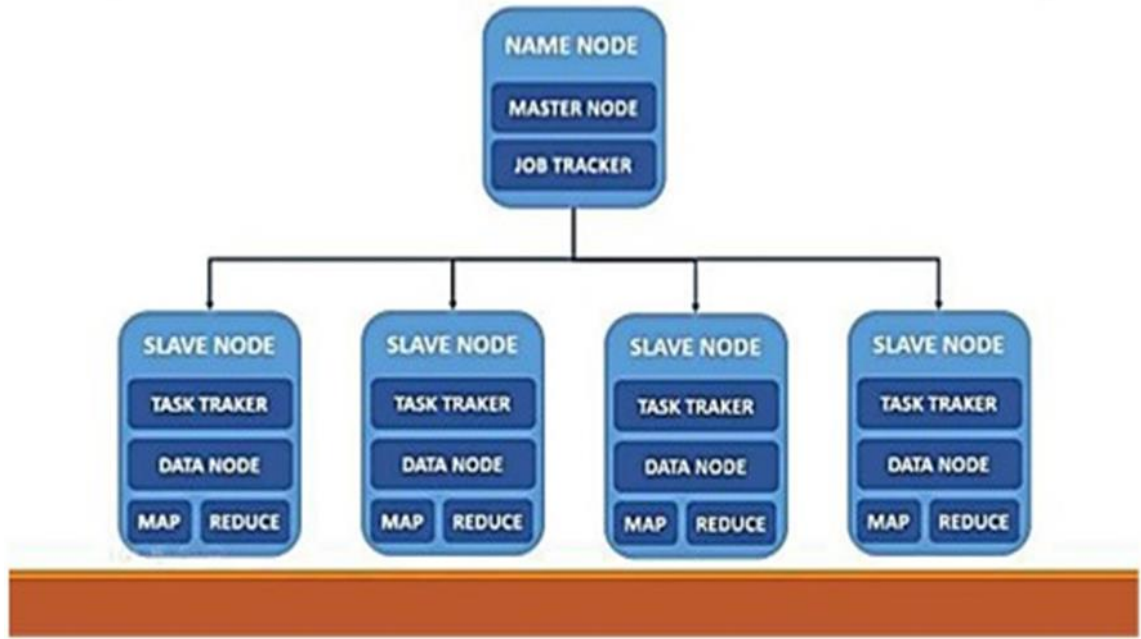

**Figure 5**. Hadoop Master/Slave Architecture

Hadoop's unravelled programming model licenses customers to quickly create and test programming in circulated structures. Performing computation on huge volumes of data has been done before, normally in a conveyed setting however composing programming for distributed frameworks is famously hard. By trading away a few programming flexibility, Hadoop makes it much simpler to write distributed programs. Because Hadoop accept practically any kind of data, it save information in far more varied formats than what is typically found in the intended rows and columns of a traditional database. Some incredible representations are machine created data and log data, worked out away arrangements including JSON, Avro and ORC. The overwhelming portion of data arranging work in Hadoop is at present being done by forming code in scripting tongues like Hive, Pig or Python.

- NameNode: handle files stored in HDFS, cited perform a function in the Map Reduce, since it is where the information will be mapped and divided into blocks, storing location information.

- DataNode: where the data actually will be stored as the objective is a dispersed application may happen several instances of a DataNode distributed in blocks.

- TaskTracker: is responsible for performing tasks or submit progress reports to the JobTracker.

- JobTracker: is responsible for managing what is being done by TaskTracker if a task fails he is responsible for relocating the task in another TaskTracker.

**6. Issues in Big Data**

The issues in Big Data are very few and while adopting the technology competently, one should clearly know by its organization. However, the Big Data issues should not be confused with harms rather these issues are important to highlight and vital to grip.

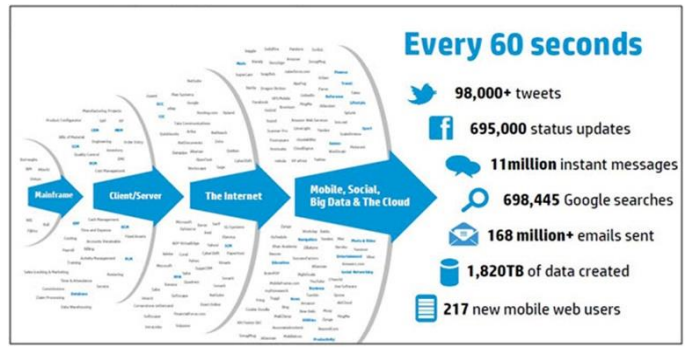

**Figure 6**. Explosion in size of Data

*6.1 Issues related to the Characteristics*

• Data Volume: As the data size increments, the estimation of different data records diminish in ratio to age, sort, riches, and amount amidst different elements. The long range interpersonal communication destinations accessible are themselves delivering information in terabytes over and over this measure of information is obviously difficult to handle with the current customary frameworks.

• Data velocity: The customary frameworks are not sufficiently skilled to play out the investigation on the information which is persistently changing or expanding. The rising online business has immediately expanded the speed and fortune of information.

• Data Variety: The data comes in various arrangements, for example, crude, organized, semi organized, and unstructured data, these distinctive configurations are difficult to handle by the reachable customary explanatory frameworks. From the investigative perspective the disappointment of customary scientific framework is a principle snag to successfully utilize the immense volume of information. In any case, jumbled data designs, unbiased data structures, and confused information semantics speak to unimportant difficulties that can prompt to investigative fall.

• Data Value: As the stored data is utilized by various associations for data analytics, which made a sort of crack between business pioneers and IT specialists, as business pioneers needs to build the estimation of their business and pick up progressively advantage not in any manner like the IT pioneers who have stress with the subtle elements of the limit and planning.

1. Storage and Transport Issues

The measure of information has exploded every time we have fanciful once more stockpiling medium. The uniqueness about the most current data explosion, primarily because of web-based social networking, is that no new storage medium has been introduced. Exabyte of data could be set up on a single PC edge work; it is unfit to clearly join the basic number of circles. Access to that data would overcome current correspondence frameworks. Tolerating that a 1 gigabyte for consistently framework has an accommodating viable conversion standard of 80%, the reasonable transmission limit is around 100 Megabytes.

2. Data Management Issues

The data information management is the most complex issue while cooking enormous information. Settling issues of perfect to use, utilize, redesigning, organization, and reference. The well springs of the information are varied by size, by setup, and by technique for social occurrence.

The management main issues are:
• Data privacy
• Security
• Ethical
• Governance.

3. Processing Issue

Acknowledge that an Exabyte of data ought to be prepared completely and arranged orderly. For simplicity, acknowledge the data is pieced into squares of 8 words, so 1 Exabyte = 1K petabytes.

Expecting a processor utilizes 100 rules on one square at 5 gigahertz, the time required for end-to-end get ready would be 20 nanoseconds. To manipulate 1K petabytes would require a total pier to pier prep time of around 232000 days. In this way, capable get ready of Exabyte of data will require wide parallel taking care of and new examination computations remembering the true objective to give fortunate and huge.

*6.2 Processing Major Issues*

- Gathering required data/information
- Arranging data from different resources (e.g., resolution when two entities are same)
- Changing the data into a form suitable for inspection
- Modelling it, whether arithmetically, or through some form of simulation
- Understanding the output, visualizing and distribution the results, think for a second how to display multifaceted analytics on an iPhone or a mobile device.

**7. Challenges in Big Data**

There are six major research areas in big data; those areas are shown in Figure 7 below.

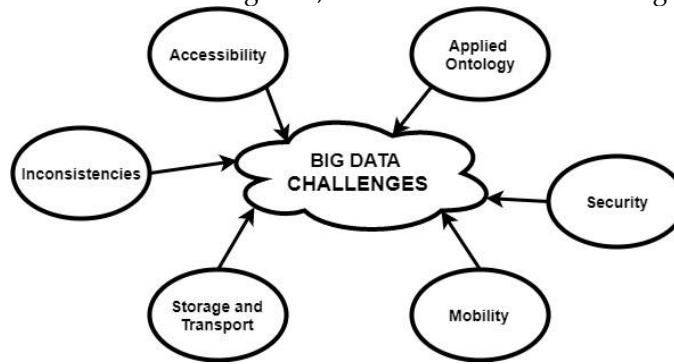

**Figure 7**. Major Research areas in Big Data

*7.1 Applied Ontology*

Applied Ontology (AO) is the method, where ontological assets to the areas, (for example, topography and bio medication) are connected. This method can be performed inside the semantic web system. Connected metaphysics analyzes the relationship between a human world and the occasions organized by them. To manufacture ranges among Semantic web, enormous data, associated data and associated reasoning. Demonstrate issues, troubles in flexibility while solidifying immense data and Semantic web. In fact, automated perspective gadgets are to be created to gather use of cosmology. Tremendous general reusable cosmology and ontological consistent frameworks to beat the planning bottleneck are the requirements of extraordinary significance.

*7.2 Security*

Big Data is worried with a considerable measure of utilization cases like Staging, Pre-handling, Processing, Meta information stockpiling and to store fleeting and long haul truth information. For serving every usage case multi-elements of establishment required. Secure and private trades are the two significant stress of IT. Be that as it may, the security and assurance transforms into a question mark as the truths volume of gigantic data rapidly create. When we think the security perspective, the accessible cryptography benchmarks cannot gather the requests of enormous information. Subsequently, security insurance is still one additionally difficult issue in enormous information.

*7.3 Storage and transport*

Big Data stores and oversee information in various courses from the customary information distribution centers. It envelops substantial sensor information; crude and semi-organized enlist information of IT businesses and the detonated measure of information from online networking. In addition, data is being made by it is conceivable that one or by all (i.e. from PDAs to supercomputers and by specialists, analyst, editorialists, researchers etc.).

*7.4 Accessibility*

The rapid pace of development in information on the web challenges the scientists to advance proficient calculations and preparing advances. The strategy for getting to enormous information has two appearances; First, the information in the source side and communicate comes about. In this way, the upgrade of scripting innovations on the program side is required to bring fundamental code from the server. Second, communicating just the genuine information in the wake of applying legitimate channels.

*7.5 Inconsistencies*

The range of this review i.e. Big Data is encompassed with multi-dimensional, specialized and precise spaces. The target of big data likewise changes over partner to partner. The Big Data Analytics is a rising edge for development and progression of innovations; in this way, its effect on society ought not to be maintained a strategic distance from. This appears to be evident that sooner the Big Data would withdraw to wrap all these areas and parts like special sciences, life and physical sciences, communiqué, capital and so forth. Big Data includes every single space; in this manner irregularity exists either in information level, data level, or learning level. This inconsistency in every level must be tended to. Irregularity has been separated in four categories i.e. temporal, textual, spatial and functional inconsistency [1].

*7.6 Mobility*

Organizations are ready to expand more resources in applications which are being acknowledged by cell phones. The incorporation of versatile processing advancements and endeavor advancement in technology has an awesome potential to swell efficiency in business settings. The heightening area based datasets, invasion of information from versatile applications, their measurement and assortment surpasses the ability of spatial and portable figuring advancements. The online conduct of various versatile clients has contributed a considerable measure to the data analytics. This union of custom controlling organizations (checking GPS and spatial data) into the gigantic data model could defy a couple of significant issues; in any case, the development in computational cost since its augmentation impacts the guiding request to PDAs, secondly, it uses geographical thinking in remotely identifying and conclusion after some time and space. The inalienable development locators in mobile phones begin a colossal measure of data from every customer's life.

## 5. Concluding Discussions

It explains on the ideas of big data took after by the applications and the difficulties confronted by it. At long last we have discussed the future open doors that could be saddle in this field. Big data is an advancing field, where a significant part of the research is yet to be finished. The data measure in all territories is detonating every day. The speed and variety of data development is expanding because of the expansion of sensor and cell phones with web association. Data produced by this way, is the best resource for enterprises in defining business procedures polices. Cloud services were utilized to prepare and break down tremendous measure of data and it has transformed into the new Big Data model to take care of the on-demand administrations. Big data at present is managed and handle of by the software named Hadoop. Be that as it may, the proliferating amount of data is making Hadoop inadequate. To outfit the capability of Big Data totally later on, broad research desires to be carried out and innovative technologies need to be developed. Big data analytics has the prospective to modify the way healthcare providers utilize advanced innovations to pick up knowledge from their clinical and other data archives and make informed decisions. Later on we'll see the quick, far reaching implementation and utilization of big data analytics over the healthcare organization and the medicinal healthcare industry.

**References**

[1] S.J.Samuel, K.RVP, K.Sashidhar, C.R.Bharathi, "A survey on big data and its research challenges", ARPN Journal of Engineering and Applied Sciences, Vol.10, No.8, Pp.3343-3347, 2015.

[2] S.Kuchipudi, T.S.Reddy, "Applications of Big data in Various Fields", International Journal of Computer Science and Information Technologies (IJCSIT), Vol.6, No.5, Pp.4629-4632, 2015.

[3] S.Mukherjee, R.Shaw, "Big Data–Concepts, Applications, Challenges and Future Scope" International Journal of Advanced Research in Computer and Communication Engineering, Vol.5, No.2, 2016.

[4] A.Misra, A.Sharma, P.Gulia, A.Bana, "Big Data: Challenges and Opportunities", International Journal of Innovative Technology and Exploring Engineering (IJITEE), Vol.4, No.2, Pp.41-42 2014.

[5] L.Venkata, S.Narayana, "A Survey on Challenges and Advantages in Big Data, International Journal of Computer Science and Technology Vol.6, No.2, 2015.

[6] H.Forest, E.Foo, D.Rose, D.Berenzon, "Big Data", white paper global transaction banking, Pp.1-26.

[7] V.Ganjir, B.K.Sarkar, R.R.Kumar, "Big data analytics for healthcare." International Journal of Research in Engineering, Technology and Science, Vol. 6, Pp.1-6, 2016.

[8] J.Sun, C.K.Reddy, "Big Data Analytics for Healthcare", Tutorial presentation at the SIAM International Conference on Data Mining Austin TX, Pp.1-112, 2013.

[9] H.C.Naik, D.Joshi, "A Hadoop Framework Require to Process Big data very easily and efficiently", International Journal of Scientific Research in Science Engineering and Technology, Vol.2, No.2, Pp.1206-1209, 2016.